\documentclass[utf8]{frontiersFPHY} 

\usepackage{url,hyperref,lineno,microtype,subcaption}
\usepackage[onehalfspacing]{setspace}

\usepackage{graphicx}
\usepackage{amsmath,amsfonts,amsthm,amssymb,bbm}
\usepackage[utf8]{inputenc}
\usepackage{color}

\newcommand{\gb}{\bar{g}}

\newcommand{\Cb}{\bar{C}}
\newcommand{\Db}{\bar{D}}
\newcommand{\Rb}{\bar{R}}

\newcommand{\cO}{{\mathcal O}}
\newcommand{\cP}{{\mathcal P}}
\newcommand{\cR}{{\mathcal R}}

\newcommand{\bgamma}{\boldsymbol{\gamma}}
\newcommand{\wt}{{w_{T}}}
\newcommand{\ws}{{w_{S}}}
\newcommand{\wtR}{{w_{T}^{\Rb}}}
\newcommand{\wsR}{{w_{S}^{\Rb}}}

\newcommand{\p}{{\partial}}

\newcommand{\be}{\begin{equation}}
\newcommand{\ee}{\end{equation}}
\newcommand{\ba}{\begin{eqnarray}}
	\newcommand{\ea}{\end{eqnarray}}
	\newcommand{\bi}{\begin{itemize}}
	\newcommand{\ei}{\end{itemize}}

\def\keyFont{\fontsize{8}{11}\helveticabold }
\def\firstAuthorLast{Saueressig {et~al.}} 
\def\Authors{Aleksandr Kurov,$^{1}$ and Frank Saueressig\,$^{2,*}$}
\def\Address{$^{1}$Theory Department, Lebedev Physics Institute, \\ Leninsky Prospect 53, Moscow 119991, Russia \\
$^{2}$Institute for Mathematics, Astrophysics and Particle Physics (IMAPP), \\ Radboud University, Heyendaalseweg 135, 6525 AJ Nijmegen, The Netherlands }
\def\corrAuthor{Frank Saueressig}

\def\corrEmail{f.saueressig@science.ru.nl}

\allowdisplaybreaks

\begin{document}
\onecolumn
\firstpage{1}

\title[Quantum Geometry of Asymptotic Safety]{On characterizing the Quantum Geometry \\ underlying Asymptotic Safety} 

\author[\firstAuthorLast ]{\Authors} 
\address{} 
\correspondance{} 

\extraAuth{}

\maketitle

\begin{abstract}
The asymptotic safety program builds on a high-energy completion of gravity based on the Reuter fixed point, a non-trivial fixed point of the gravitational renormalization group flow. At this fixed point the canonical mass-dimension of coupling constants is balanced by anomalous dimensions induced by quantum fluctuations such that the theory enjoys quantum scale invariance in the ultraviolet. The crucial role played by the quantum fluctuations suggests that the geometry associated with the fixed point exhibits non-manifold like properties. In this work, we continue the characterization of this geometry employing the composite operator formalism based on the effective average action. Explicitly, we give a relation between the anomalous dimensions of geometric operators on a background $d$-sphere and the stability matrix encoding the linearized renormalization group flow in the vicinity of the fixed point. The eigenvalue spectrum of the stability matrix is analyzed in detail and we identify a ``perturbative regime'' where the spectral properties are governed by canonical power counting. Our results recover the feature that quantum gravity fluctuations turn the (classically marginal) $R^2$-operator into a relevant one. Moreover, we find strong indications that higher-order curvature terms present in the two-point function play a crucial role in guaranteeing the predictive power of the Reuter fixed point.

\tiny
 \keyFont{ \section{Keywords:} Quantum Gravity, Asymptotic Safety, Renormalization Group, Quantum Geometry, Scaling Dimension} 
\end{abstract}

\section{Introduction}
General relativity taught us to think of gravity in terms of geometric properties of spacetime. The motion of freely falling particles is determined by the spacetime metric $g_{\mu\nu}$ which, in turn, is determined dynamically from Einstein’s equations. It is then an intriguing question what replaces the concept of a spacetime manifold once gravity is promoted to a quantum theory. Typically, the resulting geometric structure is referred to as ``quantum geometry’’ where the precise meaning of the term varies among different quantum gravity programs.

An approach towards a unified picture of the quantum gravity landscape could then build on identifying distinguished properties which characterize the underlying quantum geometry and lend themselves to a comparison between different programs. While this line of research is still in its infancy, a first step in this direction, building on the concept of generalized dimensions, has been very fruitful. In particular, the spectral dimension $d_s$, measuring the return probability of a diffusing particle in the quantum geometry, has been computed in a wide range of programs including Causal Dynamical Triangulations \citep{Ambjorn:2005db}, Asymptotic Safety \citep{Lauscher:2005qz,Reuter:2011ah,Rechenberger:2012pm,Calcagni:2013vsa}, Loop Quantum Gravity \citep{Modesto:2008jz}, string theory \citep{Atick:1988si}, causal set theory \citep{Eichhorn:2013ova,Carlip:2015mra,Eichhorn:2019uct}, the Wheeler-DeWitt equation \citep{Carlip:2009kf}, non-commutative geometry \citep{Nozari:2015iba,Kurkov:2013kfa,Alkofer:2014raa}, and Ho\v{r}ava-Lifshitz gravity \citep{Horava:2009if}, see \citep{Carlip:2017eud,Carlip:2019onx} for reviews. A striking insight originating from this comparison is that, at microscopic distances, $d_s = 2$ rather universally. The interpretation of $d_s$ as the dimension of a theories momentum space, forwarded in \citep{Amelino-Camelia:2013gna}, then suggests that the dimensional reduction of the momentum space may be a universal feature of any viable theory of quantum gravity.

Following the suggestion \citep{Pagani:2016dof},\footnote{For related ideas advocated in the context of two-dimensional gravity, see \citep{Knizhnik:1988ak,Ambjorn:1995dg}.} a refined picture of quantum geometry could use the (anomalous) scaling dimension associated with geometric operators, comprising, e.g., spacetime volumes, integrated spacetime curvatures, and geodesic distances. Within asymptotic safety program \citep{Percacci:2017fkn,Reuter:2019byg}, also reviewed in \citep{Percacci:2011fr,Litim:2011cp,Reuter:2012id,Ashtekar:2014kba,Eichhorn:2018yfc}, these quantities have been studied based on the composite operator formalism \citep{Pagani:2016dof,Becker:2018quq,Becker:2019tlf,Becker:2019fhi,Houthoff:2020zqy}. This formalism allows to determine the anomalous scaling dimension of geometric operators based on an approximation of the quantum-corrected graviton propagator.\footnote{Recently, the formalism has been generalized  to the computation of operator product expansions \citep{Pagani:2020ejb}.} For the Reuter fixed point in four dimensions the quantum corrections to the scaling of four-volumes $V_{d=4} \sim L^{4-\gamma_0}$ were determined in \citep{Pagani:2016dof}. The result $\gamma_0 = 3.986$ lent itself to the interpretation that ``spacetime could be much more empty than expected’’.
Recently, ref.\ \citep{Houthoff:2020zqy} generalized this computation by determining the anomalous scaling dimensions associated with an infinite class of geometric operators
\be\label{geoops}
\cO_n \equiv \int d^dx \sqrt{g} \, R^n \, , \qquad n = 0,1,2,\cdots \in \mathbb{N} \, , 
\ee
where $R$ denotes the Ricci scalar constructed from $g_{\mu\nu}$. While it was possible to extract analytic expressions for all $\gamma_n$, it also became apparent that the single-operator approximation underlying the computation comes with systematic uncertainties. In parallel, the anomalous scaling properties of subvolumes and geodesic distances resulting from the renormalization group fixed points underlying Stelle gravity and Weyl gravity have recently be computed in \citep{Becker:2019fhi}. In combination, the results show that the scaling of geometric quantities carries information about the renormalization group fixed point providing the high-energy completion of the theory.

The purpose of present work is two-fold: Firstly, we extend the analysis \citep{Houthoff:2020zqy} beyond the single-operator approximation and compute the complete matrix of anomalous dimensions associated with the class \eqref{geoops}. This information allows to access the spectrum of the scaling matrix. We expect that the data linked to the scaling dimensions of the geometrical operators gives a refined characterization of the quantum spacetime underlying the Reuter fixed point. Our results are closely related but complementary to the ones obtained from solving the Wetterich equation \citep{Wetterich:1992yh,Morris:1993qb,Reuter:1993kw,Reuter:1996cp} for effective average actions of $f(R)$-type \citep{Codello:2007bd,Machado:2007ea,Codello:2008vh,Benedetti:2012dx,Demmel:2012ub,Demmel:2013myx,Falls:2013bv,Demmel:2014sga,Demmel:2014hla,Falls:2014tra,Demmel:2015oqa,Dietz:2012ic,Dietz:2013sba,Dietz:2016gzg,Dietz:2015owa,Ohta:2015efa,Ohta:2015fcu,Alkofer:2018fxj,deBrito:2018jxt,Ohta:2018sze,Falls:2018ylp,Alkofer:2018baq,Burger:2019upn}. The comparison between the two complementary computations indicates that one indeed needs to go beyond the single-operator approximation in order to reconcile the results. Secondly, our work gives information on the gauge-dependence of the anomalous dimensions associated with the operators \eqref{geoops}. In this light, the value $\gamma_0 = 3.986$ found in \cite{Pagani:2016dof} may be rather extreme and quantum corrections to the scaling of volumes could be less drastic.

The rest of this work is organized as follows. Section \ref{sect.2} introduces the composite operator formalism and the propagators entering in our computation. The generating functional determining the matrix of anomalous dimensions is computed in Section \ref{sect.3}. The link to the stability matrix governing the gravitational renormalization group flow in the vicinity of the Reuter fixed point is made in Section \ref{sect.4a} and the spectral properties of the matrix are analyzed in Section \ref{sect.4b}. Section \ref{sect.5} contains our concluding remarks and comments on the possibility of developing a geometric picture of Asymptotic Safety from random geometry. The technical details underlying our computation have been relegated to three appendices: Appendix \ref{App.A} reviews the technical background for evaluating operator traces using the early-time expansion of the heat-kernel, Appendix \ref{App.B}  derives the beta functions governing the renormalization group flow of gravity in the Einstein-Hilbert truncation employing geometric gauge \citep{Benedetti:2010nr,Gies:2015tca}, and Appendix \ref{App.C} lists the two-point functions entering into the computation.

\section{Computational Framework and Setup}
\label{sect.2}
Functional renormalization group methods provide a powerful tool for investigating the appearance of quantum scale invariance and its phenomenological consequences \citep{Wetterich:2019qzx}. In particular, the Wetterich equation \citep{Wetterich:1992yh,Morris:1993qb,Reuter:1993kw,Reuter:1996cp},
\be\label{FRGE}
k \p_k \Gamma_k = \frac{1}{2} {\rm Tr} \left[ \left( \Gamma_k^{(2)} + \cR_k \right)^{-1} \,  k \p_k \cR_k \right] \, , 
\ee
plays a key role in studying the renormalization group (RG) flow of gravity and gravity-matter systems based on explicit computations. It realizes the idea of Wilson's modern viewpoint on renormalization in the sense that it captures the RG flow of a theory generated by integrating out quantum fluctuations shell-by-shell in momentum space. Concretely, eq.\ \eqref{FRGE} encodes the change of the effective average action $\Gamma_k$ when integrating out quantum fluctuations with momentum $p$ close to the coarse graining scale $k$. The flow of $\Gamma_k$ is then sourced by the right-hand side where $\Gamma_k^{(2)}$ denotes the second variation of $\Gamma_k$ with respect to the fluctuation fields, the regulator $\cR_k$ provides a $k$-dependent mass term for quantum fluctuations with momentum $p^2 \lesssim k^2$, and ${\rm Tr}$ includes a sum over all fluctuation fields and an integral over loop-momenta. Lowering $k$ ``unsuppresses'' further fluctuations which are then integrated out and change the value of the effective couplings contained in $\Gamma_k$. For later convenience, we then also introduce the ``RG-time'' $t \equiv \ln(k/k_0)$ with $k_0$ an arbitrary reference scale.

In practice, the Wetterich equation allows to extract non-perturbative information about a theories RG flow by restricting $\Gamma_k$ to a subset of all possible interaction monomials and subsequently solving eq.\ \eqref{FRGE} on this subspace. For gravity and gravity-matter systems such computations get technically involved rather quickly. Thus, it is interesting to have an alternative 
equation for studying the scaling properties of sets of operators $\cO_n$, $n=1,\cdots,N$, which are not included in $\Gamma_k$. Within the effective average action framework such an equation is provided by the composite operator equation \citep{Ellwanger:1994iz,DAttanasio:1996tzp,Litim:1998qi,Pagani:2016dof}. As a starting point, the operators $\cO_n$ are promoted to scale-dependent quantities by multiplying with a $k$-dependent matrix  $Z_{nm}(k)$ 
\be
\cO_n(k) \equiv \sum_{m}^N Z_{nm}(k) \, \cO_m \, . 
\ee
The analogy of $Z_{nm}$ to a wave-function renormalization then suggests to introduce the matrix of anomalous dimensions $\bgamma$ whose components are given by
\be\label{eq38}
\gamma_{nm} \equiv \left( Z^{-1} \p_t Z \right)_{nm} \, . 
\ee
Following the derivation \cite{Pagani:2016dof}, the $\gamma_{nm}$ can be computed from the
composite operator equation
\be\label{CompositeMaster2}
\sum_{m=1}^N \gamma_{nm} \, \cO_m = - \frac{1}{2}  {\rm Tr} \left[ \left(\Gamma_k^{(2)} + \cR_k \right)^{-1} \, \cO^{(2)}_n \, \left(\Gamma_k^{(2)} + \cR_k \right)^{-1} \, \p_t \cR_k       \right]  \, , 
\ee
where $\cO_n^{(2)}$ denotes the second functional derivative of $\cO_n$ with respect to the fluctuation fields. For the geometric operators \eqref{geoops} the evaluation of $\bgamma$ has so far focused on the diagonal matrix elements $\gamma_{nn}$, c.f.\ \cite{Pagani:2016dof,Houthoff:2020zqy}. The goal of the present work is to extend this analysis and, for the first time, study the eigenvalues of $\gamma_{ij}$ associated with the operators \eqref{geoops}.

\section{Computing the matrix of anomalous dimensions}
\label{sect.3}
The computation of  $\gamma_{nm}$ requires two inputs. First, one needs to specify the set of operators $\cO_n$. In the present work, these will be given by the geometric operators \eqref{geoops}. Secondly, one needs to specify the gravitational propagators $\Gamma_k^{(2)}$. These will be derived from $\Gamma_k$ approximated by the  Euclidean Einstein-Hilbert (EH) action
\be\label{EHaction}
\Gamma_k^{\rm EH}[g] = \frac{1}{16 \pi G_k} \int d^dx \sqrt{g} \, \left( 2 \Lambda_k - R \right) 
\ee
supplemented by a suitable choice for the gauge-fixing action \eqref{Ggf}. In practice, we obtain $\Gamma_k^{(2)}$ from the background field method, performing a linear split of the spacetime metric $g_{\mu\nu}$ into a background metric $\gb_{\mu\nu}$ and fluctuations $h_{\mu\nu}$:
\be\label{linearsplit}
g_{\mu\nu} = \gb_{\mu\nu} + h_{\mu\nu} \, . 
\ee
In order to simplify the subsequent computation, we then chose the background metric as 
 as the metric on the $d$-sphere, so that the background curvature satisfies
\be\label{background}
\Rb_{\mu\nu\rho\sigma} = \frac{\Rb}{d(d-1)} \left[\gb_{\mu\rho} \gb_{\nu\sigma} - \gb_{\mu\sigma} \gb_{\nu\rho} \right] \, , \quad 
\Rb_{\mu\nu} = \frac{\Rb}{d} \, \gb_{\mu\nu} \, , \quad 
\Db_\mu \Rb = 0 \, . 
\ee
Moreover, we carry out a transverse-traceless (TT) decomposition of the metric fluctuations \cite{York:1973ia}
\be\label{TTdec}
h_{\mu\nu} = h^{\rm T}_{\mu\nu} + \Db_\mu \xi_\nu + \Db_\nu \xi_\mu + \left( \Db_\mu \Db_\nu - \frac{1}{d} \gb_{\mu\nu} \Db^2 \right) \sigma  + \frac{1}{d} \gb_{\mu\nu} h \, , 
\ee 
where the component fields are subject to the differential constraints
\be
\gb^{\mu\nu} h^{\rm T}_{\mu\nu} = 0 \, , \qquad \Db^\mu h^{\rm T}_{\mu\nu} = 0 \, , \qquad \Db_\mu \xi^\mu = 0 \, , \qquad \gb^{\mu\nu } h_{\mu\nu} = h. 
\ee
The Jacobians associated with the decomposition \eqref{TTdec} are taken into account by a subsequent field redefinition
\be\label{fieldredef}
\sqrt{2} \left[\Delta - \frac{1}{d} \Rb \right]^{1/2} \xi_\mu \mapsto  \, \xi_\mu \, , \qquad \left[ \frac{d-1}{d} \Delta^2 - \frac{1}{d} \Rb \, \Delta \right]^{1/2} \sigma \mapsto  \sigma \, , 
\ee
and it is understood that in the sequel all propagators and the matrix elements $\cO^{(2)}_i$ are the ones associated with the rescaled fields.
In combination with the background \eqref{background}, this decomposition ensures that the differential operators appearing within the trace combine into Laplacians $\Delta \equiv - \gb^{\mu\nu} \Db_\mu \Db_\nu$ constructed from the background metric \cite{Benedetti:2010nr}.

We then specify the gauge-fixing to geometric gauge, setting $\rho = 0$ and subsequently evoking the Landau limit $\alpha \rightarrow 0$. Substituting the general form of the matrix elements listed in Table \ref{Tab.2} into the right-hand side of \eqref{CompositeMaster2} and tracing the $\alpha$-dependence one finds that the contributions of the transverse vector fluctuations $\xi_\mu$ and the scalar $\sigma$ drop out from the composite operator equation. As a consequence, the anomalous dimensions are only sourced by the transverse-traceless and conformal fluctuations. The relevant matrix elements are then readily taken from Table \ref{Tab.2}. They read
\be\label{Os}
\begin{split}
 \left. \cO_n^{(2)}\right|_{h^{\rm T}h^{\rm T}}  = & \, - \frac{1}{2} \, \Rb^{n-1} \,  \left[ n  \Delta - \left( \frac{2n(d-2)}{d(d-1)} - 1 \right) \Rb \right] \, , \\
\left. \cO_n^{(2)}\right|_{hh} \, \, = & \,  
 \frac{n(n-1) (d-1)^2}{d^2} \, \Rb^{n-2} \Delta^2
+ \frac{n(d^2 - (4n-1)(d-1)-1 )}{2d^2} \, \Rb^{n-1} \Delta \\ & \; + 
\left( \frac{d-2}{4d} - \frac{n(d-n-1)}{d^2} \right) \Rb^n  \, , 	
\end{split}
\ee
together with
\be\label{Gammas}
\begin{split}
\Gamma^{(2)}_k|_{h^{\rm T}h^{\rm T}}  = & \, \frac{1}{32 \pi G_k} \left[\Delta - 2 \Lambda_k + C_T \Rb \right]  \, ,  \\	
\Gamma^{(2)}_k|_{hh} \, \, = & \, - \frac{(d-1)(d-2)}{32 \pi G_k \, d^2} \left[ \Delta - \frac{d}{d-1} \Lambda_k + C_S \Rb \right] \, ,
\end{split}
\ee
and
\be\label{Rreg2}
\begin{split}
	\big. \cR_k \big|_{h^{\rm T}h^{\rm T}} = & \, \frac{1}{32 \pi G_k} \, R_k \, , \qquad 
	\big. \cR_k \big|_{hh} =  - \frac{(d-1)(d-2)}{32 \pi G_k \,  d^2} \, R_k \, .
\end{split}
\ee
Here
\be\label{Cdef}
C_T \equiv \frac{d^2-3d+4}{d(d-1)} \, , \quad C_S \equiv \frac{d-4}{2(d-1)} \, , 
\ee
and $R_k(\Delta) = k^2 \, r(\Delta/k^2) $ is a scalar regulator function which later on will be specified to the Litim regulator \eqref{ropt}.

Substituting the expressions \eqref{Os}-\eqref{Rreg2} into the composite operator equation \eqref{CompositeMaster2} then yields
\be\label{anom2}
\gamma_{nm} = - \frac{1}{2} \Big( {\rm Tr}_{T} \left[ \, W_{T}(n;\Delta) \, \right]   +  {\rm Tr}_{S} \left[ \, W_{S}(n;\Delta) \, \right] \Big) \Big|_{\cO_m} \, . 
\ee
Here the subscripts ${T}$ and ${S}$ indicate that the trace is over transverse-traceless ($T$) and scalar ($S$) fluctuations, respectively, an the symbol $|_{\cO_m}$ indicates the projection of the right-hand side onto the operator $\cO_m$. The explicit form of the operator-valued functions $W_{T}$ and $W_{S}$ is 
\be\label{Wfcts}
\begin{split}
	W_{T}(n;\Delta) \equiv & \, 16 \pi G_k \left[ P_k - 2 \Lambda_k + C_{\rm T} \,  \Rb\right]^{-2}	\, \Rb^{n-1} 
	\left[ - n  \Delta + \left( \tfrac{2n(d-2)}{d(d-1)} - 1 \right) \Rb \right]
	\left(\p_t R_k - \eta_N R_k \right) \, , \\
	W_{S}(n;\Delta) \equiv & -  \frac{16 \pi G_k}{(d-2)} \left[ P_k - \frac{d}{d-1} \Lambda_k + C_{h} \Rb\right]^{-2} \, \Rb^{n-2} \, \\ & \; 	
	\left[ 2n(n-1)(d-1) \,  \Delta^2
	+ n(d + 2 - 4n) \, \Rb \Delta + 
	\tfrac{d^2 - 2 d (2 n+1) + 4 n (n+1) }{2(d-1)} \Rb^2 \right]
	\left(\p_t R_k - \eta_N R_k \right) \, ,
\end{split}
\ee

Before delving into the explicit evaluation of the traces, the following structural remark is in order. Inspecting \eqref{Wfcts}, one observes that the right-hand side associated with the $n$th row contains at least $\Rb^{n-2}$ powers of the background curvature. This entails that the matrix of anomalous dimensions has the following triangular form
\be\label{gammamat}
\bgamma = \left[
\begin{array}{cccccccc}
	\gamma_{00} & {\color{blue}{\gamma_{01}}} & {\color{blue}{\gamma_{02}}} & {\color{blue}{\gamma_{03}}} & {\color{blue}{\gamma_{04}}} & {\color{blue}{\gamma_{05}}} & {\color{blue}{\gamma_{06}}} & {\color{blue}{\cdots}} \\
		\gamma_{10} & {\gamma_{11}} & {\color{blue}{\gamma_{12}}} & {\color{blue}{\gamma_{13}}} & {\color{blue}{\gamma_{14}}}& {\color{blue}{\gamma_{15}}} & {\color{blue}{\gamma_{16}}} & {\color{blue}{\cdots}} \\
			\gamma_{20} & {\gamma_{21}} & \gamma_{22} & {\color{blue}{\gamma_{23}}} & {\color{blue}{\gamma_{24}}} & {\color{blue}{\gamma_{25}}} & {\color{blue}{\gamma_{26}}} & {\color{blue}{\cdots}} \\
			0 & {\gamma_{31}} & \gamma_{32} & \gamma_{33} & {\color{blue}{\gamma_{34}}} & {\color{blue}{\gamma_{35}}} & {\color{blue}{\gamma_{36}}}  & {\color{blue}{\cdots}} \\
			0 & 0 & \gamma_{42} & \gamma_{43} & \gamma_{44} & {\color{blue}{\gamma_{45}}} & {\color{blue}{\gamma_{46}}}  & {\color{blue}{\cdots}} \\
			0 & 0 & 0 & \gamma_{53} & \gamma_{54} & \gamma_{55} & {\color{blue}{\gamma_{56}}}  & {\color{blue}{\cdots}} \\
\end{array}
\right] \, . 
\ee 
The explicit value of the matrix entries \eqref{anom2} is readily computed employing the heat-kernel techniques reviewed in Appendix \ref{App.A}. In practice, we will truncated the heat-kernel expansion at order $R^2$, setting the coefficients $a_n, n \ge 3$ to zero. This is in the spirit of the
``paramagnetic approximation'' suggested in \cite{Nink:2012vd}, that the curvature terms relevant for asymptotic safety originate from the curvature terms contained in the propagators. For the matrix entries $\gamma_{nm}$ this entails that all entries on the diagonal and below (marked in black) are computed exactly while contributions to the terms above the diagonal (marked in blue) will receive additional contributions from higher-orders in the heat-kernel. In particular all entries $\gamma_{nm}$ with $m \ge n+3$ are generated solely from expanding the curvature terms proportional to $C_T$ and $C_S$ in the transverse-traceless and scalar propagators.

Evaluating \eqref{anom2} based on these approximations then results in an infinite family of generating functionals $\Gamma_n(\Rb), n \ge 0 \in \mathbb{N}$:
\be\label{eq:gen}
\begin{split}
	\Gamma_n(\Rb) = & \, \frac{16 \pi g}{(4 \pi)^{d/2}} \Big[
	c_1^T \, q^2_{d/2+1}(\wtR) \left(\tfrac{\Rb}{k^2}\right)^{-1} 
	+ c_2^T\,  q^2_{d/2}(\wtR) \\ & \; \qquad \qquad
	+ c_3^T\,  q^2_{d/2-1}(\wtR) \left(\tfrac{\Rb}{k^2}\right) 
	+ c_4^T\, q^2_{d/2-2}(\wtR) \left(\tfrac{\Rb}{k^2}\right)^2 
	\\ & \; \qquad \qquad
	+ c_1^S \, q^2_{d/2+2}(\wsR) \left(\tfrac{\Rb}{k^2}\right)^{-2} 
	+ c_2^S \, q^2_{d/2+1}(\wsR ) \left(\tfrac{\Rb}{k^2}\right)^{-1}
	+ c_3^S \, q^2_{d/2}(\wsR)
	\\ & \; \qquad \qquad
	+ c_4^S \, q^2_{d/2-1}(\wsR) \left(\tfrac{\Rb}{k^2}\right) 
	+ c_5^S \, q^2_{d/2-2}(\wsR) \left(\tfrac{\Rb}{k^2}\right)^2  
	\Big] \, . 
\end{split}
\ee
Here we introduced the dimensionless couplings
\be\label{dimless}
g_k = k^{d-2} \, G_k \, , \qquad \lambda_k = \Lambda_k \, k^{-2} \, ,  
\ee
and the anomalous dimension of Newton's coupling $\eta_N \equiv (G_k)^{-1} \p_t G_k$. The threshold functions $q^p_n(w)$ are defined in eq.\ \eqref{qfcts} and their arguments in the transverse-traceless and scalar sector are
\be\label{defarg}
\wt = - 2 \lambda \, , \qquad \ws \equiv - \frac{d}{d-1} \lambda
\, , \qquad 
\wtR = \wt + C_{T} \Rb/k^2\, , \qquad \wsR \equiv \ws + C_{S} \Rb/k^2
\, .
\ee
The coefficients $c_k^i$ depend on $d$ and $n$. In the tensor sector they are given by
\be
\begin{array}{ll}
	c_1^T =  \, \frac{1}{2}  n \, d \, a_0^T \, , \qquad & 
	c_2^T =  \frac{1}{2} n(d-2) a_1^T - \left( \frac{2n(d-2)}{d(d-1)}-1 \right) \, a_0^T \, , \\[1.3ex]
	c_3^T =  \frac{1}{2} n(d-4) a_2^T - \left( \frac{2n(d-2)}{d(d-1)}-1 \right) \, a_1^T \, , \qquad \qquad & 
	c_4^T =   - \left( \frac{2n(d-2)}{d(d-1)}-1 \right) \, a_2^T \, .
\end{array}
\ee
Their counterparts in the scalar sector read
\be
\begin{split}
c_1^S = &  \, \frac{n(n-1)(d-1)d(d+2)}{2(d-2)} \, a_0^S , \\
c_2^S = &  \,  \frac{1}{2} n  (n-1)\, d(d-1) \, a_1^S + \frac{nd(d+2-4n)}{2(d-2)} \, a_0^S  \, , \\
c_3^S = & \,  \frac{1}{2}  n (n-1) \, (d-1)(d-4) \, a_2^S +  \frac{1}{2}  n(d+2-4n) \, a_1^S +  \frac{d^2-2d(2n+1)+4n(n+1)}{2(d-1)(d-2)} \, a_0^S \, , \\
c_4^S = & \,	 \frac{ n(d-4)(d+2-4n)}{2(d-2)} \, a_2^S +  \frac{d^2-2d(2n+1)+4n(n+1)}{2(d-1)(d-2)} \, a_1^S  \, , \\
c_5^S = & \, \frac{d^2-2d(2n+1)+4n(n+1)}{2(d-1)(d-2)} \, a_2^S \, . 
\end{split}
\ee
Finally, the $a_n^i$ are the heat-kernel coefficients listed in Table \ref{Tab.1}. 

The entries in $\bgamma$ are then generated as the coefficients of the Laurent series expansion
\be\label{gen:series}
\Gamma_n(\Rb)  = \sum_{m=-2}^\infty \, \gamma_{n,n+m} \, \Rb^m \, , \qquad n \ge 0, \; m+n \ge 0 \, . 
\ee
For instance, the two lines of entries below the diagonal, $\gamma_{n,n-2}, n \ge 2$ and $\gamma_{n,n-1}, n \ge 1$, obtained in this way are
\be\label{eq:gnn}
\begin{split}
\gamma_{n,n-2} = & \, \frac{16 \pi g}{(4 \pi)^{d/2}} \, \frac{n(n-1)(d-1)d(d+2)}{2(d-2)} \, k^4 \,  q^{2}_{d/2+2}(\ws) \, , \\
\gamma_{n,n-1} = & \frac{16 \pi g}{(4 \pi)^{d/2}} \, n \, d \, k^2 \, \Big[ 
\frac{1}{4} (d-2)(d+1) \, q^{2}_{d/2+1}(\wt)
+ \frac{1}{12} (n-1)(d-1)  q^{2}_{d/2+1}(\ws) \\ & \qquad \qquad 
+ \frac{d+2-4n}{2(d-2)} \,  q^{2}_{d/2+1}(\ws)
- \frac{(n-1)(d-4)(d+2)}{2(d-2)}  q^{3}_{d/2+2}(\ws)
\Big] \, . 
\end{split}
\ee
Eqs.\ \eqref{eq:gen} - \eqref{gen:series} constitute the main result of this work. They give completely analytic terms \emph{for all entries of the anomalous dimension matrix} $\bgamma$.   

At this stage, a few remarks are in order. 
\bi
\item[{\bf 1)}] The entries of the anomalous dimension matrix carry a specific $k$-dependence: $\gamma_{nm} \propto (k^2)^{n-m}$. This can be understood by noticing that the matrix $\bgamma$ acts on operators $\cO_m$ with different canonical mass dimensions. The $k$-dependence then guarantees that the eigenvalues of $\bgamma$ are  independent of $k$.
\item[{\bf 2)}] The entries $\gamma_{n,n-2}$ are solely generated from the scalar contributions, i.e., the transverse-traceless fluctuations do not enter into these matrix elements. Technically, this feature is associated with the Hessians $\cO^{(2)}_n$, cf.\ Table \ref{Tab.2}: the matrix elements in the scalar sector start at $\Rb^{n-2}$ while the transverse-traceless sector starts at $\Rb^{n-1}$. 
\item[{\bf 3)}] Notably, $d=4$ is special. In this case the entries above the diagonal, $\gamma_{nm}$ with $m \ge n+3$ are generated from the transverse-traceless sector only. All contributions from the scalar sector are proportional to at least one power of $C_{S}$ and thus vanish if $d=4$.
\item[{\bf 4)}] The matrix $\bgamma$ is a function of the (dimensionless) couplings entering the Einstein-Hilbert action. Thus $\bgamma$ assigns a set of anomalous dimensions to every point in the $g$-$\lambda$--plane. Since $\bgamma$ is proportional to $g$, the magnitude of the anomalous dimensions becomes small if $g \ll 1$. In particular, $\bgamma$ vanishes at the Gaussian fixed point $g_* = \lambda_* = 0$ where one recovers the classical scaling of the geometric operators. 
\ei
\section{Scaling analysis for the Reuter fixed point}
\label{sect.4}
Starting from the general result \eqref{gen:series}, we now proceed and discuss its implications for the quantum geometry associated with Asymptotic Safety.
\subsection{Relating the scaling of geometric operators and the RG flow}
\label{sect.4a}
By construction, the matrix $\bgamma$ assigns anomalous scaling dimensions to any point in the $g$-$\lambda$ plane. In order to characterize the quantum geometry related to Asymptotic Safety, we study the properties of this matrix at the Reuter fixed point found in Appendix \ref{App.B}, cf.\ eq.\ \eqref{ReuterFPpos}
\be\label{rfp1}
\begin{split}
	& \mbox{Reuter fixed point:} \quad d=3: \qquad g_* = 0.198 \, , \quad \lambda_* = 0.042 \, , \quad \lambda_* g_*^2 =  1.65 \times 10^{-3} \, ,  \\
	& \mbox{Reuter fixed point:} \quad d=4: \qquad g_* = 0.911 \, ,  \quad \lambda_* = 0.160 \, , \quad \lambda_* g_* =  0.146 \, .
\end{split} 
\ee
From the definition of the beta function $\p_t u_n = \beta_{u_n}({u_i})$ and the fact that at a fixed point $\beta_{u_n}({u_i^*}) = 0$, it follows that the properties of the RG flow in the vicinity of the fixed point are encoded in the stability matrix ${\bf B} = [B_{nm}]$, 
\be\label{linflow}
\p_t u_n(k) = \sum_m {B}_{nm} (u_m(k) - u_m^*) \, , \qquad {B}_{nm} \equiv \left. \frac{\partial \beta_{u_m}}{\partial u_n} \right|_{u=u^*} \, . 
\ee
Let us denote the eigenvalues of ${\bf B}$ by $\lambda_n$ so that spec(${\bf B}$) = $\{\lambda_n\}$. Eq.\ \eqref{linflow} then entails that eigendirections corresponding to eigenvalues with a negative (positive) real part attract (repel) the RG flow when $k$ is increased, i.e., they correspond to UV-relevant (UV-irrelevant) directions. The number of UV-relevant directions then gives the number of free parameters which are not fixed by the asymptotic safety condition: along these directions the RG flow automatically approaches the Reuter fixed point as $k\rightarrow \infty$.

Formally, one can then derive a relation between $\bgamma$ and the stability matrix $\bf B$ \cite{Pagani:2016pad,Houthoff:2020zqy},
\be\label{Bmatdef}
B_{nm} = - d_n \delta_{nm} + \gamma_{nm} \, , 
\ee
where $d_n = d - 2n$ is the canonical scaling dimension of the operator $\cO_n$. This relation is remarkable in the following sense: The construction of the (approximate) fixed point solution \eqref{rfp1} is based on the two operators $\cO_0$ and $\cO_1$, comprising the Einstein-Hilbert truncation. The relation \eqref{Bmatdef} then shows that \emph{the matrix of anomalous dimensions carries information about the stability properties of the Reuter fixed point beyond the set of operators which are considered when solving the Wetterich equation to locate the fixed point}.
 We illustrate this idea by studying the spectrum of $B_{nm}$ obtained at the fixed points \eqref{rfp1}. Before embarking on this discussion, the following cautious remark is in order though. While the composite operator formalism may allow to obtain information on the stability properties of a fixed point beyond the approximation used for the propagators, it is also conceivable that the formalism becomes unreliable for eigenvalues $\lambda_n$ with $n \ge N_{\rm max}$. Heuristically, this is suggested by the following argument: when studying fixed point solutions in the $f(R)$-approximation the propagators  include powers of $\Rb$ beyond the linear terms captured by the Einstein-Hilbert action. These terms give rise to additional contributions in the generating functional \eqref{gen:series} which may become increasingly important in assessing the spectrum of ${\bf B}$ for eigenvalues with increasing numbers of $n$. This picture is also suggested by our results in Section \ref{sect.4b}.

This said, we now investigate the properties of the stability matrix \eqref{Bmatdef}. Here we will resort to the following frameworks:
\bi
\item[{\bf I}] The spectrum of of ${\bf B}$ generated by the full generating functional \eqref{eq:gen} including the contribution of zero-modes in the heat-kernel for $d=4$.
\item[{\bf II}] In the conformally reduced approximation \citep{Reuter:2008wj}. In this case, the contribution of the tensor fluctuations is set to zero by hand, so that $\bgamma$ contains the contribution from the scalar trace in \eqref{anom2} only.
\ei
The latter choice is motivated by the observation that this framework gives rise to the spec({\bf B}) which is the most robust under increasing the size of the matrix ${\bf B}$. Clearly, one could easily envision other approximations which could be applied to the general result \eqref{eq:gen}. Examples include the exclusion of the zero-mode terms appearing in $d=4$ or the  ``sparse approximation'' where only two lines above and below the diagonal are non-trivial, i.e., the entries in the upper-triangular sector which are solely created by expanding the curvature terms contained in the gravitational propagators are eliminated. In order to understand the working (and limitations) of the conformal operator formalism, the frameworks {\bf I} and {\bf II} are sufficient though. We checked by explicit computations that the exclusion of zero-modes or evaluating the spectrum of ${\bf B}$ in the sparse approximation leads to the same qualitative picture.

\subsection{Spectral properties of the stability matrix}
\label{sect.4b}
We first give the diagonal entries $\gamma_{nn}$ within framework {\bf I}. This corresponds to the ``single-operator approximation'' of the composite operator formalism employed in \citep{Pagani:2016dof,Houthoff:2020zqy}. At the fixed points \eqref{rfp1} one finds
\be\label{diagentry}
\begin{split}
& d= 3: \qquad \gamma_{nn}^* = 0.653 - 0.872 n - 0.029 n^2 \, , \\ 
& d= 4: \qquad \gamma_{nn}^* = 2.299 - 3.765 n \, .
\end{split}
\ee	
These relations exhibit two remarkable features. Firstly, the structure of $\cO^{(2)}_n$ (cf.\ Table \ref{Tab.2}) entails that the entries of $\bgamma$ are second order polynomials in $n$. It is then remarkable that the diagonal entries essentially follow a linear scaling law up to $n \approx 30$ ($d=3$) or even exactly ($d=4$). Secondly, eq.\ \eqref{diagentry} entails that the diagonal entries of the stability matrix {\bf B} \emph{are always negative}. Thus the single-operator approximation predicts that all eigendirections of the Reuter fixed point in the $f(R)$-space are UV-attractive. It was noted in \cite{Houthoff:2020zqy} that this is actually in tension with results obtained from solving the Wetterich equation on the same space. On this basis, it is expected that the off-diagonal entries in $\bgamma$ play a crucial role in determining the spectrum of $\bf B$.

We now discuss the properties of the stability matrices $\bf B$ evaluated at the Reuter fixed points \eqref{rfp1}. The generating functional \eqref{eq:gen} allows to generate truncations of $\bf B$  of size $N=100$ rather easily and determine the resulting spectrum of eigenvalues numerically. The structure of $\bf B$ then entails that there is always one eigenvalue which is independent of the matrix size. For framework {\bf I} its value is given by
\be\label{universalev}
\begin{split}
& d=3: \qquad \lambda_1^{\bf \rm I} = -2.347\, , \\ 
& d=4: \qquad \lambda_1^{\bf \rm I} = -1.701 \, .
\end{split}
\ee
In the conformally reduced approximation (framework {\bf II}) in $d=3$ this feature extends to the second eigenvalue as well
\be
d=3: \qquad \lambda_2^{\bf \rm II} = -2.828 \, , \qquad  \lambda_2^{\bf \rm II} = -0.967 \, . 
\ee

\begin{figure}[t]
	\begin{center}
\includegraphics[width=5.5cm]{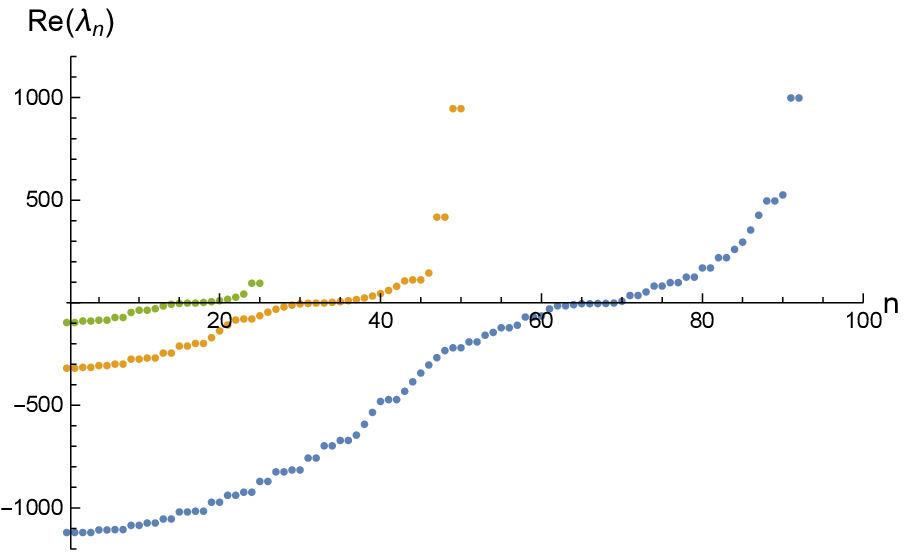} \,
\includegraphics[width=5.5cm]{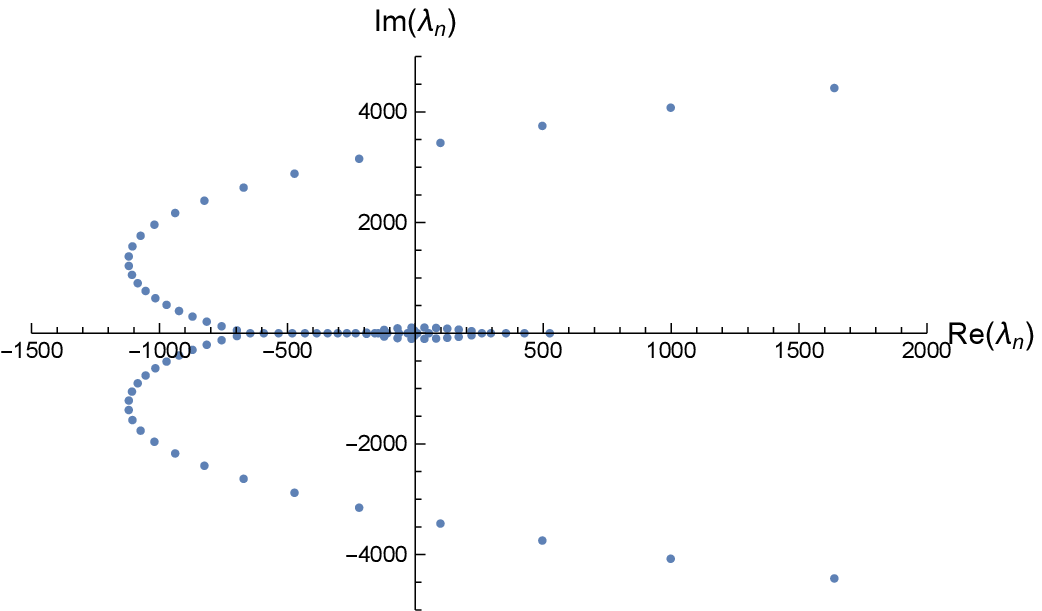} \,
\includegraphics[width=5.5cm]{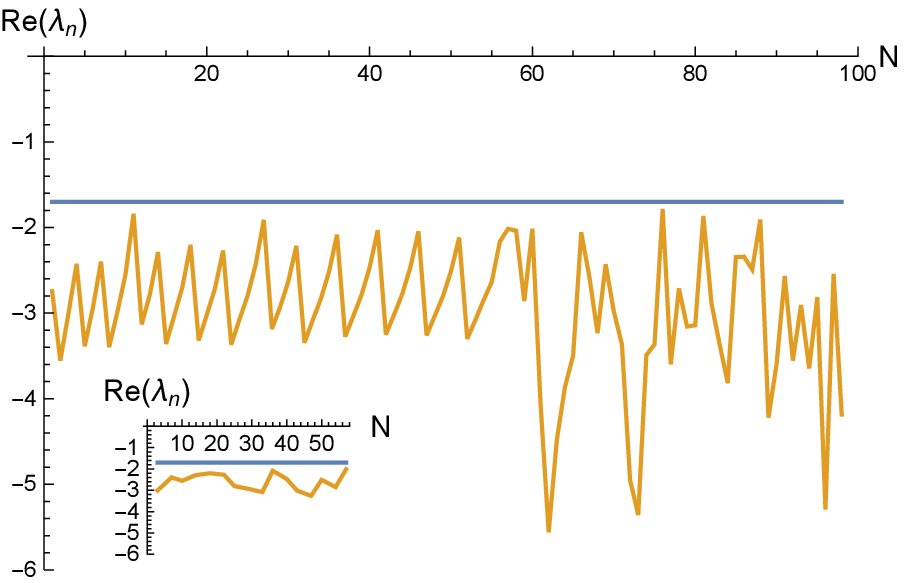} 
\end{center}
\caption{\label{Fig.spectrum1} Spec(${\bf B}$) in $d=4$ dimensions obtained within framework {\bf I}. The left diagram displays the real parts Re$(\lambda_n)$ of the eigenvalues found for the stability matrices of sizes $N=25$ (left line, green dots), $N=50$ (middle line, orange dots) and $N=100$ (right line, blue dots). The middle diagram shows the location of the eigenvalues $\lambda_n$ ($N=100$) in the complex plane. The right diagram traces the value of the first two relevant eigenvalues as a function of the matrix size $N$.}
\end{figure}

\begin{figure}[t]
	\begin{center}
		\includegraphics[width=5.5cm]{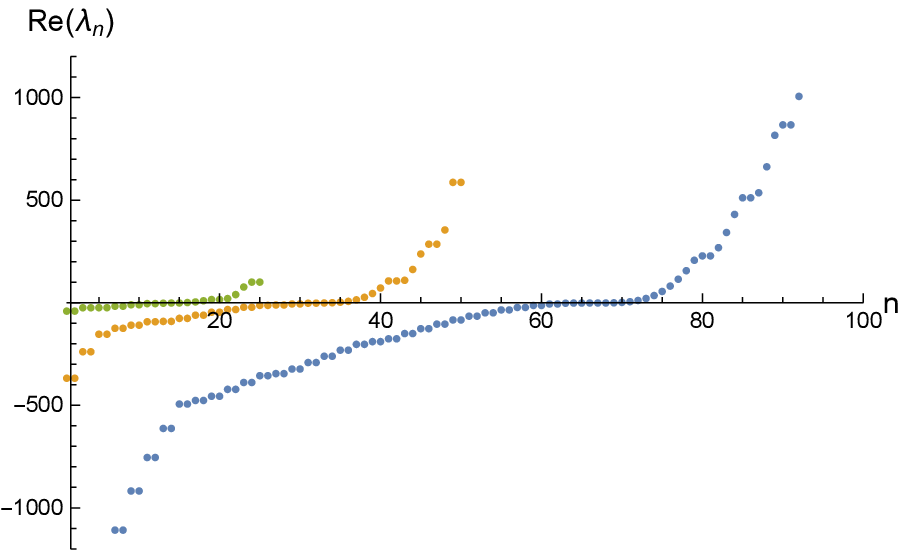} \,
		\includegraphics[width=5.5cm]{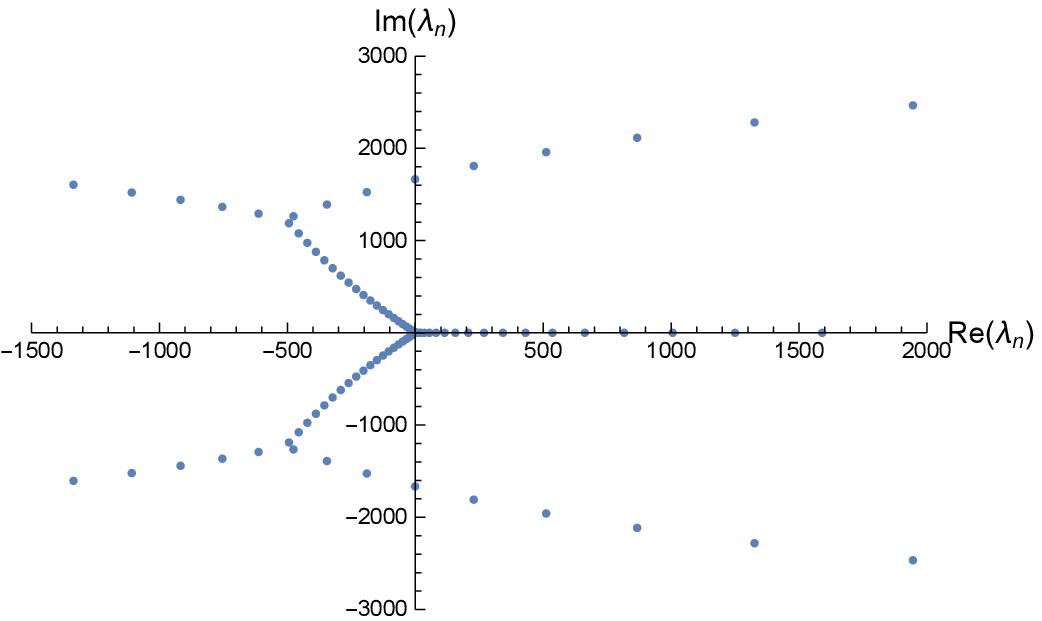} \,
		\includegraphics[width=5.5cm]{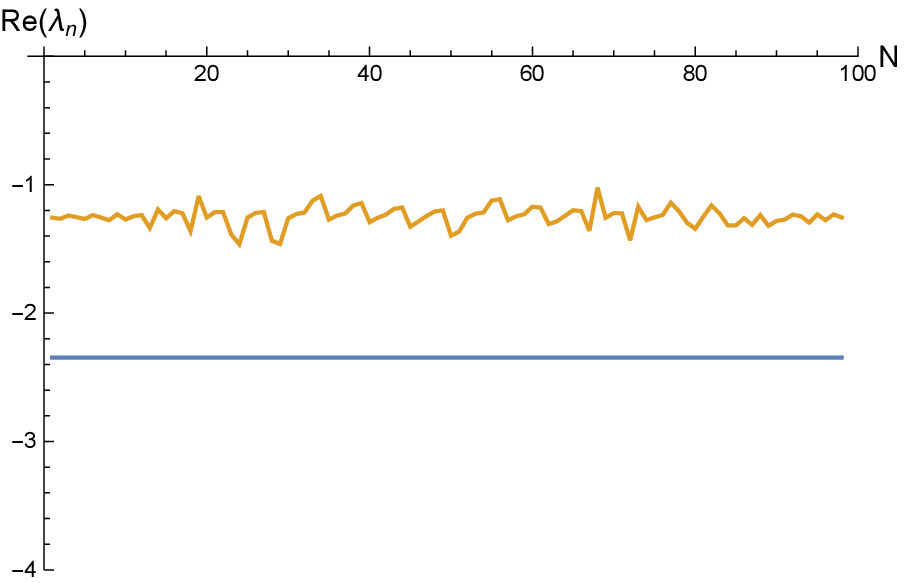} 
	\end{center}
	\caption{\label{Fig.spectrum2} Spec(${\bf B}$) in $d=3$ dimensions obtained within framework {\bf I}. The left diagram displays the real parts Re$(\lambda_n)$ of the eigenvalues found for the stability matrices of sizes $N=25$ (left line, green dots), $N=50$ (middle line, orange dots) and $N=100$ (right line, blue dots). The middle diagram shows the location of the eigenvalues $\lambda_n$ ($N=100$) in the complex plane. The right diagram traces the value of the first two relevant eigenvalues as a function of the matrix size $N$.}
\end{figure}

\begin{figure}[t]
	\begin{center}
		\includegraphics[width=5.5cm]{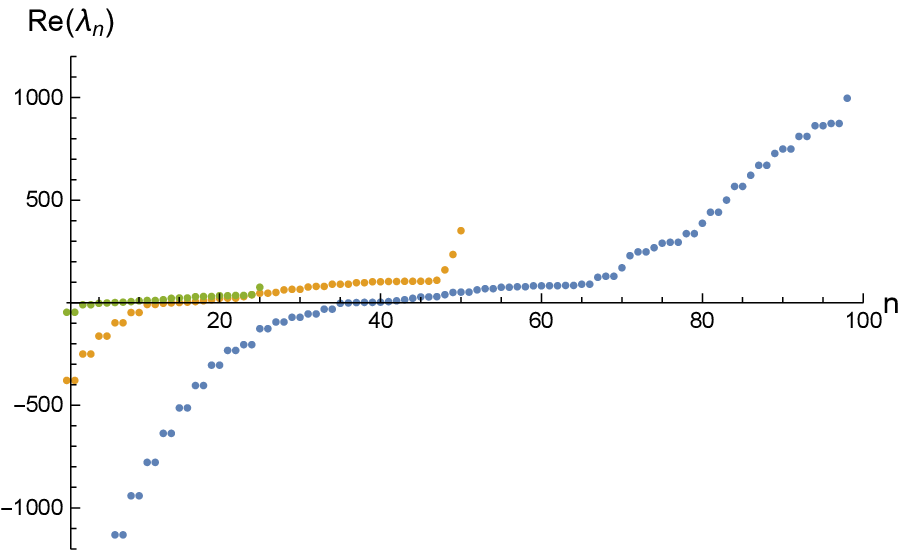} \,
		\includegraphics[width=5.5cm]{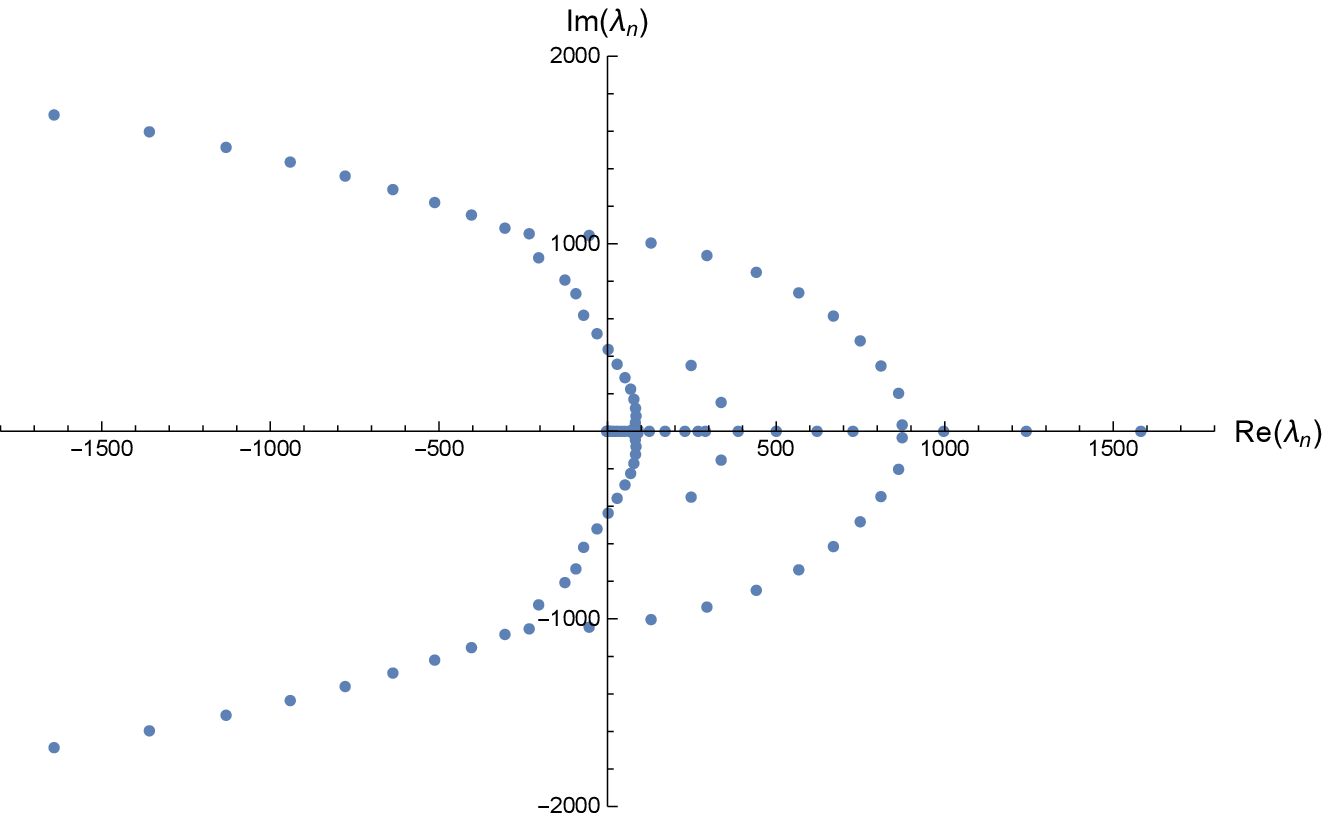} \,
		\includegraphics[width=5.5cm]{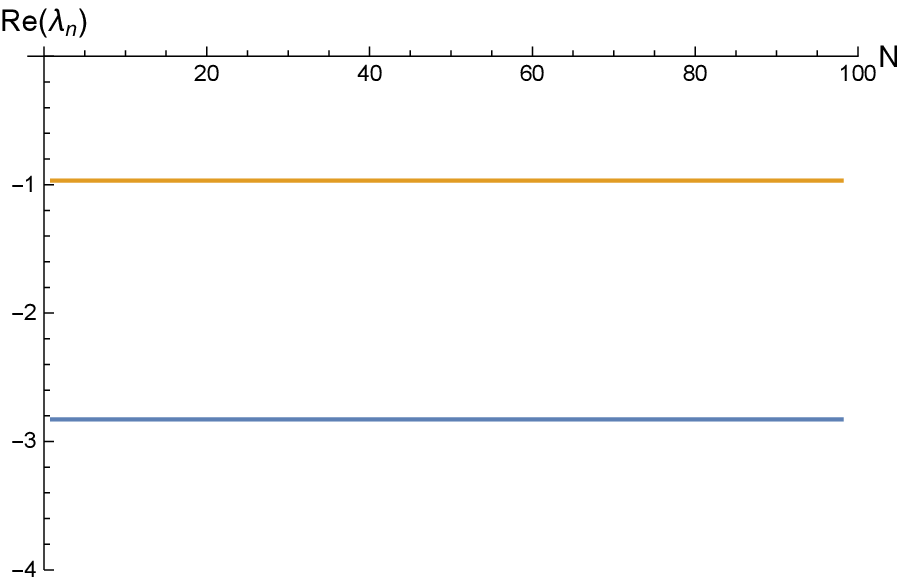} 
	\end{center}
	\caption{\label{Fig.spectrum3} Spec(${\bf B}$) in $d=3$ dimensions obtained within framework {\bf II}. The left diagram displays the real parts Re$(\lambda_n)$ of the eigenvalues found for the stability matrices of sizes $N=25$ (left line, green dots), $N=50$ (middle line, orange dots) and $N=100$ (right line, blue dots). The middle diagram shows the location of the eigenvalues $\lambda_n$ ($N=100$)in the complex plane. The right diagram traces the value of the first two relevant eigenvalues as a function of the matrix size $N$.}
\end{figure}

The properties of spec({\bf B}) beyond these universal eigenvalues obtained from the framework {\bf I} in $d=4$ and $d=3$ as well as in the conformally reduced approximation in $d=3$ (framework {\bf II}) are shown in Figs.\ \ref{Fig.spectrum1}, \ref{Fig.spectrum2} and \ref{Fig.spectrum3}, respectively.
The left diagrams show the real part, Re$(\lambda_n)$, $n=1,\cdots,N$ of the stability matrices of size $N=25$ (left line, green dots), $N=50$ (middle line, orange dots), and $N=100$ (right line, blue dots). The lines clearly illustrate that increasing $N$ adds additional eigenvalues coming with both increasingly positive and increasingly negative real parts. This feature is shared by all frameworks discussed above. The middle diagrams illustrate the location of spec({\bf B}) for $N=100$ in the complex plane. While the patterns are quite distinct, they share the existence of nodes where complex eigenvalues are created which then move out into the complex plane along distinguished lines. The right diagrams trace the first two negative eigenvalues as a function of the matrix size $N$. In all cases, the structure of $\bf B$ implies that the first eigenvalue is independent of $N$ while the other parts of the spectrum exhibit an $N$-dependence. As illustrated in Figs.\ \ref{Fig.spectrum1} and \ref{Fig.spectrum2}, the eigenvalues $\lambda_n$, $n \ge 2$ follow intriguing periodicity patterns. The average over the second and third eigenvalues found in the matrices of size up to $N=100$ (for $\bar{\lambda}_2$) and $N=20$ (for $\bar{\lambda}_3$, excluding values where a complex eigenvalue has appeared in the interval spanned by $\lambda_1$ and $\lambda_3$) are\footnote{Our errors are purely statistical, giving the standard deviation based on the data set of eigenvalues. An estimate of the systematic errors is highly non-trivial and will not be attempted in this work.}
\be\label{specmean}
\begin{split}
& d=3: \qquad \bar{\lambda}_2^{\bf \rm I} = -1.25 \pm 0.08  \, , \qquad  \bar{\lambda}_3^{\bf \rm I}  = -0.61 \pm 0.40 \\
& d=4: \qquad  \bar{\lambda}_2^{\bf \rm I} = -2.86 \pm 0.61 \, , \qquad  \bar{\lambda}_3^{\bf \rm I}  = -6.36 \pm 2.04  \, .  \\
\end{split}
\ee 

Carefully analyzing the $N$-dependence of spec({\bf B}) reveals that there is a close relation between the distribution of eigenvalues in the complex plane (middle diagrams) and the oscillations of $\lambda_2$ visible in the left diagrams: the oscillations are linked to the appearance of new complex pairs of eigenvalues. Focusing  on the four-dimensional case where this feature is most prominent, one finds that singling out the values of $\lambda_2$ \emph{just before the occurrence of the new pair of complex eigenvalues in} spec({\bf B}) essentially selects the $\lambda_2(N)$ constituting the maxima in the oscillations. The resulting subset of eigenvalues is displayed in the inset shown in Fig.\ \ref{Fig.spectrum1} and is significantly more stable than the full set. The statistical analysis shows that in this case
\be
d=4: \qquad \bar{\lambda}_2^{{\bf I},{\rm subset}} = -2.61 \pm 0.39 \, , 
\ee 
so that the fluctuations are reduced by a factor two as compared to the full set \eqref{specmean}.

At this stage, it is interesting to compare the averages \eqref{specmean} to the eigenvalue spectrum obtained from the smallest non-trivial stability matrix $\bf B$ with size $N=3$:
\be\label{specN3}
\begin{split}
	& d=3: \qquad \lambda_2^{\bf \rm I} = -2.35  \, , \qquad  \lambda_2^{\bf \rm I} = -1.26  \, , \qquad  \lambda_3^{\bf \rm I}  = -0.20 \, ,  \\
	& d=4: \qquad   \lambda_2^{\bf \rm I} = -1.70 \, , \qquad \lambda_2^{\bf \rm I} = -2.74 \, , \qquad  \lambda_3^{\bf \rm I}  =-5.95   \, .  \\
\end{split}
\ee 
Thus we conclude that small values of $N$ already give a good estimate of the (averaged) spectrum of $\bf B$.

We close this section with a general remark on the structure of spec($\bf B$). The stability matrix is not tied to the Reuter fixed point but well-defined on the entire $g$-$\lambda$--plane: the generating functional \eqref{eq:gen} assigns an infinite tower of eigenvalues to each point in this plane. At the Gaussian fixed point, $(\lambda_*,g_*) = (0,0)$, $\bgamma = 0$ and spec($\bf B$) follows from classical power counting. The strength of the quantum corrections to spec($\bf B$) is then controlled by the values of $g$ and $\lambda$. In particular, there is a region in the vicinity of the Gaussian fixed point where these corrections are small. This motivates defining ``perturbative domains'' $\cP$ by the condition that spec($\bf B$) is dominated by its classical part. Concretely, we define
\be\label{def:pertdomain}
\begin{split}
& d=3: \qquad \left\{ 
\begin{array}{l}
	\cP_2 = \{ (\lambda,g) | \, {\rm spec}({\bf B}) \, \mbox{has 2 UV-relevant eigenvalues} \} \\[1.2ex]
	\cP_3 = \{ (\lambda,g) | \, {\rm spec}({\bf B}) \, \mbox{has 3 UV-relevant eigenvalues} \} 
\end{array}
\right.	\\[1.3ex]
& d=4: \qquad \quad \; \cP_3 = \{ (\lambda,g) | {\rm spec}({\bf B}) \, \mbox{has 3 UV-relevant eigenvalues} \} \, . 
\end{split}
\ee
Loosely speaking, the definitions of these domains corresponds to imposing that the quantum corrections are not strong enough to turn more than one classically UV-marginal ($d=4$) or UV-irrelevant ($d=3$) eigendirection into a relevant one. 

\begin{figure}[t]
	\begin{center}
		\includegraphics[width=8.3cm]{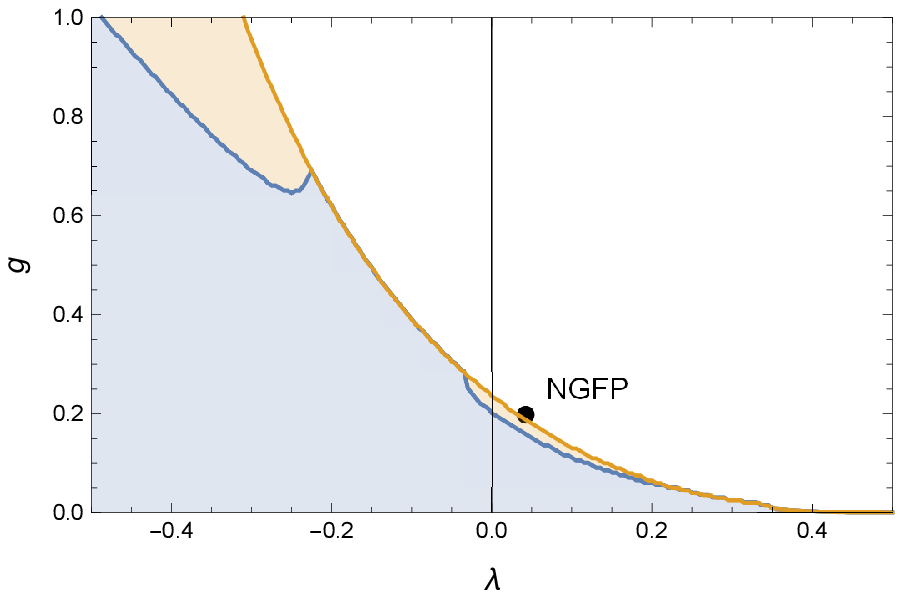} \; \;
		\includegraphics[width=8.3cm]{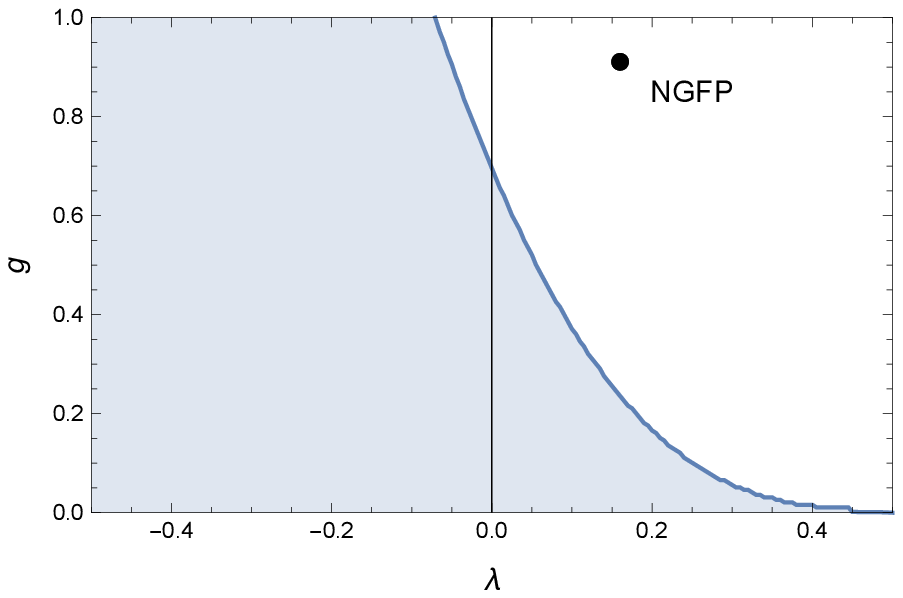}
	\end{center}
	\caption{\label{Fig.perturbative} Spectral analysis for the matrices $\bf B$ of size $N=10$ as a function of $g$ and $\lambda$ in $d=3$ (left diagram) and $d=4$ (right diagram). In the shaded region spec({\bf B}) is dominated by its classical part. In $d=3$ the blue and orange regions support two and three negative eigenvalues, respectively, while in $d=4$ the blue region supports three negative eigenvalues. The boundary to the white region is set by the appearance of a new, complex pair of eigenvalues coming with a negative real part. The Reuter fixed points \eqref{rfp1} are marked by the black dots and are located outside the shaded regions.}
\end{figure}

Fig.\ \ref{Fig.perturbative} illustrates the shape of the domains $\cP$ obtained from the spectrum of the stability matrices with $N=10$ (framework $\bf I$) in $d=3$ (left panel) and $d=4$ (right panel). 
In $d=3$ the regions $\cP_2$ and $\cP_3$ are shaded in blue and orange, respectively while in $d=4$ $\cP_3$ is shaded blue. At the boundary of these regions a new complex pair of eigenvalues with negative real part appears in the spectrum which then violates the definitions \eqref{def:pertdomain}. Within the present computation the Reuter fixed points \eqref{rfp1} are located outside of $\cP_3$ which is consistent with the eigenvalue spectra shown in Figs.\ \ref{Fig.spectrum1} and \ref{Fig.spectrum2}.  
%

\section{Conclusions and Outlook}
\label{sect.5}
In this work, we applied to composite operator formalism to construct a completely analytic expression for the matrix $\bgamma$ encoding the anomalous scaling dimensions of the geometrical operators $\cO_n \equiv \int d^dx \sqrt{g} R^n$, $n \in \mathbb{N}$, on a background sphere. Our work constitutes the first instance where 
the composite operator formalism for gravity is extended beyond the single-operator approximation.
 Within the geometric gauge adopted in our work, the anomalous dimensions originate from the transverse-traceless and trace mode of the gravitational fluctuations. The gauge-modes, corresponding to the vector sector of the transverse-traceless decomposition, decouple. Our derivation made two assumptions: firstly, we assumed that the propagators of the fluctuation fields can be approximated by the (gauge-fixed) Einstein-Hilbert action. Secondly, we assumed that terms appearing in the early-time expansion of the heat-kernel beyond the $R^2$-level can be neglected. On this basis, we derived the generating functional \eqref{eq:gen} from which the matrix of anomalous dimensions \eqref{gammamat} can be generated efficiently.

 As illustrated in Section \ref{sect.4} the stability matrix $\bf B$ resulting from the composite operator formalism allows to study the stability properties of the Reuter fixed point. This novel type of analysis provided the following structural insights on Asymptotic Safety:
 \bi
 \item[{\bf 1)}] The composite operator approach suggests that in $d=4$ quantum fluctuations turn the classically marginal $R^2$-operator into a UV-relevant one. Similarly, the analysis in $d=3$ dimensions predicts that the classically irrelevant $R^2$-coupling becomes UV-relevant. 
 \item[{\bf 2)}] The eigenvectors of ${\bf B}$ do not coincide with the geometric operators $\cO_n$. In general they are given by linear combinations containing an infinite number of terms.
 \item[{\bf 3)}] The non-diagonal terms $\gamma_{nm}, n \not = m$ play a crucial role in determining the spectrum of ${\bf B}$. Within the assumptions made in our derivation one furthermore finds that  increasing the size of ${\bf B}$ creates complex pairs of eigenvalues which wander through the complex plain and lead to new (most likely spurious) UV-relevant directions.  
 \ei

The analysis of the spectrum of the stability matrix as a function of the dimensionless Newton coupling $g$ and cosmological constant $\lambda$ reveals the existence of a domain where the eigenvalues are dominated by classical power counting. The resulting spectrum is then similar to the one encountered when solving the Wetterich equation in the polynomial $f(R)$-approximation which determined the eigenvalues of the stability matrix for $N=6$ \cite{Codello:2007bd,Machado:2007ea}, $N=8$ \cite{Codello:2008vh}, $N=35$ \cite{Falls:2013bv,Falls:2014tra}, and lately also $N=71$ \cite{Falls:2018ylp}. In particular, ref.\ \cite{Falls:2018ylp} reported that for large values of $n$ the real parts of the eigenvalues $\lambda_n$ follow an almost Gaussian behavior

\be\label{fRfit}
\lambda_n^{f(R)} \approx  a \, n - b  \, , \qquad a = 2.042 \pm 0.002 \, , \;  b = 2.91 \pm 0.05 \, .
\ee
where $a$ and $b$ are the best-fit values.  As indicated in Figure \ref{Fig.perturbative}, the present computation places the Reuter fixed point outside of this scaling domain, i.e., for sufficiently large matrices one obtains new eigenvalues coming with both positive and negative real parts. This makes it conceivable that the higher-order curvature terms appearing in the propagators of the $f(R)$-approximation play a crucial role in extending the domain such that it includes the fixed point, thereby guaranteeing its predictive power. Conversely, one may use the structure of the stability matrix to analyze the conditions on its entries such that its eigenvalues exhibit ``apparent convergence'' discussed in \citep{Denz:2016qks}.

Arguably, the most intriguing result of our work is the spectral analysis of the stability matrix showing the distributions of its eigenvalues in the complex plane, c.f.\ the middle diagrams of Figs.\ \ref{Fig.spectrum1}, \ref{Fig.spectrum2}, and \ref{Fig.spectrum3}. The resulting patterns are reminiscent of the Lee-Yang theory for phase transitions \citep{Blythe_2003}. This suggests two immediate applications. First, the status of Asymptotic Safety makes it conceivable that there are actually an infinite number of Reuter-type fixed points arising from gravity and gravity-matter systems. Understanding the characteristic features of their eigenvalue distributions in terms of nodal points creating complex eigenvalues may then constitute a powerful tool for classifying these fixed points and giving a precise definition to the notion of ``gravity-dominated’’ renormalization group fixed points in gravity-matter systems. Secondly, tracing the eigenvalues $\lambda_n$ along their Lee-Yang type orbits in the complex plane could provide a novel tool for testing the convergence of the eigenvalue distribution of ${\bf B}$ also outside of the perturbative domains \eqref{def:pertdomain} where the spectrum is governed by classical power counting. Clearly, it would be interesting to follow up on these points in the future.

As a by-product our analysis also computed the diagonal entries of the anomalous dimension matrix in geometric gauge, cf.\ eq.\ \eqref{diagentry}. It is instructive to compare this result to the value of the diagonal entries obtained in harmonic gauge \citep{Pagani:2016dof,Houthoff:2020zqy}
\be
\begin{split}
	& d= 3: \qquad \gamma_{nn}^* = 1.591 - 1.505n - 0.118n^2 \, , \\ 
	& d= 4: \qquad \gamma_{nn}^* = 3.987 - 4.733n - 0.095n^2 \, .
\end{split}
\ee	
This identifies two features which are robust under a change of gauge-fixing: in both cases, the values of $\gamma_{nn}$ up to $n \simeq \cO(10)$ follows a linear scaling law: in all cases the coefficients multiplying the quadratic terms are small or even vanishing when adopting geometric gauge in four dimensions. Secondly, the entries in the stability matrix $B_{nn}$ are negative definite for all values $n$. At the same time, this comparison gives a first idea of the accuracy to which the composite operator formalism in the single-operator approximation is capable to determine the anomalous scaling dimension of the geometric operators: most likely, the results have the status of order-of-magnitude estimates: they should not be interpreted as ``precision results'' which one should try and reproduce to the given accuracy. Conceptually, it would be interesting to understand (and eliminate) the gauge-dependence of the result. Most likely, this will require imposing on-shell conditions to the master equation \eqref{CompositeMaster2}, following e.g., the ideas outlined in \citep{Benedetti:2011ct,Falls:2015qga}. We leave this point to future work though.

As one of its most intricate features, the composite operator formalism employed in this work could act as a connector between Asymptotic Safety \citep{Percacci:2017fkn,Reuter:2019byg} and more geometric approaches to quantum gravity based on causal dynamical triangulations \citep{Ambjorn:2012jv,Loll:2019rdj} or random geometry. In $d=2$ dimensions, a natural benchmark would involve a quantitative comparison of scaling properties associated with the geodesic length recently considered in \citep{Pagani:2016dof,Becker:2018quq,Becker:2019tlf,Becker:2019fhi,Houthoff:2020zqy} and exact computations for random discrete surfaces in the absence of matter
fields \citep{Ambjorn:1995dg,Le_Gall_2007} as well as rigorous and numerical bounds arising from
Liouville Gravity in the presence of matter \citep{Ding:2018uez,Barkley:2019kvp}. On the renormalization group side this will involve taking limits akin to \citep{Nink:2015lmq}. Conversely, it is interesting to generalize the two-dimensional constructions to higher dimensions. The connection between the stability matrix ${\bf B}$ and the anomalous scaling dimension $\bgamma$ of geometric operators may then be an interesting link allowing to probe Asymptotic Safety based on geometric constructions of a quantum spacetime.

\newpage
\appendix
\section*{Appendices}
\renewcommand{\thesection}{\Alph{section}}
\section{Heat-kernel, Mellin Transforms, and Threshold functions}
\label{App.A}
\begin{table}[t]
	\begin{center}
		\begin{tabular}{cccc}
			& $S$ &  $TV$ & $T$ \\ \hline \hline
			$\bigg. a_0^i$ & $1$  & $d-1$ & $\frac{(d-2)(d+1)}{2}$ \\
			$\bigg. a_1^i$ & $\frac{1}{6}$  & \quad $\frac{(d+2)(d-3)}{6d}$ \quad & \quad $\frac{(d+1)(d+2)(d-5+3 \delta_{d,2})}{12(d-1)}$ \quad \\
			$\bigg. a_2^i$ & \quad $\frac{5d^2-7d+6}{360 d (d-1)}$ \quad & $-$ & $\frac{(d+1)(5d^4-22d^3-83d^2-392d-228 + 1440 \delta_{d,2} + 3240 \delta_{d,4})}{720 d (d-1)^2}$\\ \hline \hline 
		\end{tabular}
		\caption{\label{Tab.1} Heat-kernel coefficients $a_n^i$ for scalars ($S$), transverse vectors ($TV$), and transverse-traceless symmetric tensors ($T$) on a background $d$-sphere \cite{Lauscher:2001ya}. The terms proportional to $\delta_{d,2}$ and $\delta_{d,4}$ are linked to zero modes of the decomposition \eqref{TTdec} on the $2$- and $4$-sphere. The dash $--$ indicates that the corresponding coefficient is not entering into the present computation.}
	\end{center}
\end{table}
The calculation of $\bgamma$ requires the evaluation of the operator traces appearing on the right-hand side of the composite operator equation \eqref{CompositeMaster2}. This computation can be done effectively by applying the early-time heat-kernel expansion for minimal second-order differential operators $\Delta \equiv - \gb^{\mu\nu} \Db_\mu \Db_\nu$. Following the ideas advocated in \cite{Lauscher:2001ya,Benedetti:2010nr}, we carry out a transverse-traceless decomposition of the  fluctuation fields. Paired with a maximally symmetric background geometry, this decomposition ensures that all differential operators in the trace arguments organize themselves into Laplacians $\Delta$.

These traces can then be evaluated using the Seeley-deWitt expansion of the heat-kernel on the $d$-sphere $S^d$:
\be\label{heatkernel}
\left. {\rm Tr}_i \left[e^{-s\Delta}\right] \right|_{S^d}  \simeq  \frac{1}{(4\pi s)^{d/2}} \int d^dx \sqrt{\gb} \left[ a_0^i + a_1^i \, s \, \Rb + a_2^i \, s^2 \, \Rb^2 + \ldots\right] \, .
\ee
Here $i =\{{S,TV,T}\}$ labels the type of field on which the Laplacian acts and the dots represent higher-order curvature terms. The relevant coefficients $a_n^i$ have been computed in \cite{Lauscher:2001ya} and are listed in Table \ref{Tab.1}. Their derivation manifestly uses the identities \eqref{background} in order to simplify the heat-kernel expansion on a general manifold \cite{Vassilevich:2003xt}.

The expansion \eqref{heatkernel} is readily generalized to functions of the Laplacian. Introducing the $Q$-functionals
\be
Q_n[W] \equiv \frac{1}{\Gamma(n)} \int_0^\infty dz z^{n-1} W(z) \, , \; \; n > 0 \, , \qquad \quad Q_0[W] = W(0) \, ,
\ee
one has \cite{Codello:2008vh}
\be\label{heat2}
\begin{split}
	{\rm Tr}_i\left[W(\Delta) \right] = & \, \frac{1}{(4\pi)^{d/2}} \int d^dx \sqrt{\gb} \, \big[ a_0^i \, Q_{d/2}[W] + a_1^i \, Q_{d/2-1}[W] \, \Rb  
	 + a_2^i \, Q_{d/2-2}[W] \, \Rb^2 + \ldots \big] \, . 
\end{split}
\ee  

In order to write $\bgamma$ and the beta functions of the Einstein-Hilbert truncation in a compact form, it is convenient to express the $Q$-functionals in terms of the dimensionless threshold functions \cite{Reuter:1996cp}
\be\label{threshold}
\begin{split}
	\Phi^p_n(w) \equiv & \, \frac{1}{\Gamma(n)} \int_0^\infty dz \, z^{n-1} \, \frac{r(z) - z r^\prime(z)}{[z + r(z) + w]^p} \, , \\
	\widetilde{\Phi}^p_n(w) \equiv & \, \frac{1}{\Gamma(n)} \int_0^\infty dz \, z^{n-1} \, \frac{r(z)}{[z + r(z) + w]^p} \, . 
\end{split}
\ee
Here $r(z)$ is the dimensionless profile function associated with the scalar regulator $R_k(z) = k^2 \, r(z)$ introduced in eq.\ \eqref{regdef} and the prime denotes a derivative with respect to the argument. For later convenience we also define the combination
\be\label{qfcts}
q^p_n(w) \equiv \Phi^p_n(w) - \frac{1}{2} \, \eta_N \, \widetilde{\Phi}^p_n(w) \, . 
\ee 

The arguments of the traces appearing in $\bgamma$, eq.\ \eqref{anom2}, and the Einstein-Hilbert truncation studied in Appendix \ref{App.B} have a canoncial form. Defining $P_k \equiv z + R_k(z)$, the identity 
\be\label{Q2map}
Q_n\left[z^q \, (P_k + w k^2)^{-p} \, G_k \p_t \left( G_k^{-1} R_k \right) \right] = 2 \, \frac{\Gamma(n+q)}{\Gamma(n)} \, (k^2)^{n+q+1-p} \, q^p_{n+q}(w) 
\ee
 allows to convert the corresponding $Q$-functionals into the dimensionless threshold functions. For $q=0$ this reduces to
\be\label{Qphimap}
\begin{split}
	& \,  Q_n\left[ (P_k + w k^2)^{-p} \, \p_t R_k \right] = 2 \, (k^2)^{n+1-p} \, \Phi^p_n(w) \, , \\ 
	& \, Q_n\left[ (P_k + w k^2)^{-p} \, G_k \p_t \left( G_k^{-1} R_k \right) \right] = 2 \, (k^2)^{n+1-p} \, q^p_n(w) \, . 
\end{split}
\ee
Notably, the second set of identities suffices to derive the beta functions of the Einstein-Hilbert truncation while the evaluation of $\bgamma$ requires the generalization \eqref{Q2map}.

For maximally symmetric backgrounds the background curvature $\Rb$ is covariantly constant. As a consequence, it has the status of a parameter and can be included in the argument of the threshold functions. Expansions in powers of $\Rb$ can then be constructed from the recursion relations
\be\label{phirec}
\frac{d}{dw} \, \Phi^p_n(w) = - p \, \Phi^{p+1}_n(w) \, , \qquad
\frac{d}{dw} \, q^p_n(w) = - p \, q^{p+1}_n(w) \, . 
\ee

Throughout the work, we specify the (scalar) regulator 
\be\label{regdef}
R_k(\Delta) = k^2 \, r(\Delta/k^2) \, , 
\ee
to the Litim regulator \cite{Litim:2000ci,Litim:2001up}. In this case the dimensionless profile function $r(z)$ is given by 
\be\label{ropt}
r(z) = (1-z) \Theta(1-z) \, , 
\ee
with $\Theta(x)$ the unit-step function. For this choice the integrals \eqref{threshold} can be carried out analytically, yielding
\be\label{thresholdLitim}
\Phi^{p, {\rm Litim}}_n(w) = \frac{1}{\Gamma(n+1)} \, \frac{1}{(1+w)^p} \, , \quad
\widetilde{\Phi}^{p, {\rm Litim}}_n(w) = \frac{1}{\Gamma(n+2)} \, \frac{1}{(1+w)^p}\, . 
\ee
%

\section{The Einstein-Hilbert truncation in general gauge}
\label{App.B}
Structurally, the composite operator equation provides a map from the couplings contained in the Hessian $\Gamma_k^{(2)}$ to the matrix of anomalous dimensions $\bgamma$. This map is independent of the RG flow entailed by the Wetterich equation. In order to characterize the geometry associated with the Reuter fixed point, the map has to be evaluated at the location of the fixed point. This appendix then studies the flow of $\Gamma_k$ in the Einstein-Hilbert truncation supplemented by a general gauge-fixing term. The key result is the position of the Reuter fixed point, eq.\ \eqref{rfp1}, which underlies the spectral analysis of Sect.\ \ref{sect.4}. Our analysis essentially follows \cite{Lauscher:2001cq,Benedetti:2010nr,Gies:2015tca},  to which we refer for further details. 

The Einstein-Hilbert truncation approximates the effective average action $\Gamma_k[h;\gb]$ by the Einstein-Hilbert action $\Gamma_k^{\rm EH}[g] = \frac{1}{16 \pi G_k} \int d^dx \sqrt{g} \, \left( 2 \Lambda_k - R \right)$ supplemented by a gauge-fixing functional $\Gamma_k^{\rm gf}[h;\gb]$ and the corresponding ghost action $S^{\rm ghost}[h,\bar{C},C;\gb]$
\be\label{EHtrunc}
\Gamma_k[h;\gb] \simeq \Gamma_k^{\rm EH}[g] + \Gamma_k^{\rm gf}[h;\gb] + S^{\rm ghost}[h,\bar{C},C;\gb] \, . 
\ee
This ansatz contains two scale-dependent couplings, Newton's coupling $G_k$ and the cosmological constant $\Lambda_k$. In the present analysis, we work with a generic gauge-fixing term

\be\label{Ggf}
 \Gamma_k^{\rm gf}[h;\gb] = \frac{1}{32 \pi G_k \, \alpha} \int d^dx \sqrt{\gb} \gb^{\mu\nu} F_\mu F_\nu \, , \qquad F_\mu \equiv \Db^\nu h_{\mu\nu} - \frac{1+\rho}{d} \Db_\mu h \, , 
\ee
where $\alpha$ and $\rho$ are free, dimensionless parameters. The harmonic gauge used in \cite{Pagani:2016pad,Houthoff:2020zqy} corresponds to $\alpha = 1, \rho = d/2-1$ while the present computation significantly simplifies when adopting geometric gauge, setting $\rho=0$ before evoking the Landau limit $\alpha \rightarrow 0$. The ghost action associated with \eqref{Ggf} is
\be\label{Sghost}
S^{\rm ghost}[h,\bar{C},C;\gb] = - \int d^dx \sqrt{\gb} \, \bar{C}_\mu \left[\Db^\rho \delta^\mu_\nu D_\rho + \Db^\rho g_{\rho\nu} D^\mu - \frac{2(1+\rho)}{d} \Db_\mu \gb^{\sigma\rho} g_{\rho\nu} D_\sigma  \right] C^\nu
\ee
Following the strategy employed in the gravitational sector, c.f.\ eq.\ \eqref{TTdec}, the fields $\Cb_\mu$, $C^\mu$ are decomposed into their transverse and longitudinal parts
\be\label{Tdec}
C^\mu = C^{{\rm T}\mu} + \Db^\mu \, \eta \, , \qquad \Db_\mu C^{{\rm T}\mu} = 0 \, , 
\ee
followed by a rescaling 
\be\label{fredghost}
\Delta^{1/2} \, \eta \mapsto \eta \, .
\ee
 The part of the ghost action quadratic in the fluctuation fields then becomes
\be\label{ghostquad}
S^{\rm ghost,quad} = \int d^dx \sqrt{\gb} \left\{ 
\bar{C}^{\rm T}_\mu \left[ \Delta - \frac{1}{d} \Rb \right] C^{{\rm T}\mu} + 2 \bar{\eta} \left[ \frac{d-1-\beta}{d} \Delta - \frac{1}{d} \Rb \right] \eta 
\right\} \, . 
\ee

We now proceed by constructing the non-zero entries of the Hessian $\Gamma^{(2)}_k$. These are obtained by expanding $\Gamma_k$ to second order in the fluctuation fields, substituting the transverse traceless decomposition \eqref{TTdec} and \eqref{Tdec}, and implementing the field redefinitions \eqref{fieldredef} and \eqref{fredghost}. Subsequently taking two functional variations with respect to the fluctuation fields then leads to the matrix elements listed in the middle block of Table \ref{Tab.2}.

The final ingredient entering the right-hand side of the Wetterich equation is the regulator $\cR_k$. We generate this matrix from the substitution rule 
\be\label{type1reg}
\Delta \mapsto P_k \equiv \Delta + R_k(\Delta) \, , 
\ee
dressing each Laplacian by a scalar regulator $R_k(\Delta)$. The latter then provides 
a mass for fluctuation modes with momentum $p^2 \lesssim k^2$. In the nomenclature introduced in \cite{Codello:2008vh} this corresponds to choosing a type I regulator.  The non-zero entries of  $\cR_k$ generated in this way are listed in the bottom block of Table \ref{Tab.2}.

We now have all the ingredients to compute the beta functions resulting from the Wetterich equation projected onto the Einstein-Hilbert action. Adopting the geometric gauge $\rho = 0, \alpha \rightarrow 0$ used in the main section, all traces appearing in the equation simplify to the $Q$-functionals evaluated in eq.\ \eqref{Qphimap}. Defining
\be\label{betaEH}
 \p_t g_k = \beta_g(g_,\lambda_k;d) \equiv (d-2+\eta_N) g_k \, , \qquad  \p_t \lambda_k \equiv \beta_\lambda(g_,\lambda_k;d)
\ee
where the anomalous dimension of Newton's coupling is parameterized by \cite{Reuter:1996cp}
\be
\eta_N(g,\lambda) = \frac{g B_1(\lambda)}{1-g B_2(\lambda)} \, , 
\ee
the explicit computation yields
\be
\beta_\lambda = (\eta_N - 2) \lambda + \frac{g}{(4\pi)^{d/2-1}} \left[
(d-2)(d+1) q^1_{d/2}(\wt) + 2 q^1_{d/2}(\ws) + 2 d q^1_{d/2}(0) - 4d \Phi^1_{d/2}(0)
\right] \, , 
\ee
and
\be
\begin{split}
B_1 =  & \, \frac{1}{(4\pi)^{d/2-1}} \Big[ 
\frac{(d+1)(d+2)(d-5)}{3(d-1)} \Phi^1_{d/2-1}(\wt) 
- \frac{2(d-2)(d+1)(d^2-3d+4)}{d(d-1)} \Phi^2_{d/2}(\wt) \\ & \, 
+ \frac{2}{3} \Phi^1_{d/2-1}(\ws) - \frac{2(d-4)}{d-1} \Phi^2_{d/2}(\ws) 
- \frac{4(d^2-d+1)}{d(d-1)}\Phi^2_{d/2}(0) - \frac{2(d^2-6)}{3d} \Phi^1_{d/2-1}(0)
\Big] \, , \\
B_2 = & \, - \frac{1}{2(4\pi)^{d/2-1}} \Big[
\frac{(d+1)(d+2)(d-5)}{3(d-1)} \widetilde{\Phi}^1_{d/2-1}(\wt) 
- \frac{2(d-2)(d+1)(d^2-3d+4)}{d(d-1)} \widetilde{\Phi}^2_{d/2}(\wt) \\ & \,  
+ \frac{2}{3} \widetilde{\Phi}^1_{d/2-1}(\ws) - \frac{2(d-4)}{d-1} \widetilde{\Phi}^2_{d/2}(\ws) + \frac{4(d^2-d+1)}{d(d-1)}\widetilde{\Phi}^2_{d/2}(0) +\frac{2(d^2-6)}{3d} \widetilde{\Phi}^1_{d/2-1}(0) 
\Big] \, . \\
\end{split}
\ee
Here the threshold functions $\Phi^p_n$, $\widetilde{\Phi}^p_n$ and $q^p_n(w)$ are defined in eqs.\ \eqref{threshold} and \eqref{qfcts} and their arguments $\wt$ and $\ws$ have been introduced in \eqref{defarg}.

It is now straightforward to localize the Reuter fixed point by determining the roots of the beta functions \eqref{betaEH} numerically. For the Litim regulator \eqref{ropt} this yields
\be\label{ReuterFPpos}
\begin{split}
& \mbox{Reuter fixed point:} \quad d=3: \qquad g_* = 0.198 \, , \quad \lambda_* = 0.042 \, , \\
& \mbox{Reuter fixed point:} \quad d=4: \qquad g_* = 0.911 \, ,  \quad \lambda_* = 0.160 \, . \\
\end{split}
\ee
Analyzing the stability properties of the RG flow in its vicinity, it is found that the fixed point constitutes a UV attractor, with the eigenvalues of the stability matrix given by
\be\label{ReuterStab}
\begin{split}
& \mbox{Reuter fixed point:} \quad d=3: \qquad \lambda^{\rm EH}_{1,2} = -1.658 \pm 0.546 i \, ,  \\
& \mbox{Reuter fixed point:} \quad d=4: \qquad \lambda^{\rm EH}_{1,2} = -2.132 \pm 2.697 i \, . \\
\end{split}
\ee
These results agree with the ones found in \cite{Benedetti:2010nr} at the 10\% level. The difference can be traced back to the two distinct regularization procedures employed in the computations, so that the findings are in qualitative agreement. This completes our analysis of the Einstein-Hilbert truncation underlying the scaling analysis in the main part of this work.
\section{Matrix-elements of geometric operators}
\label{App.C}
The expansions of $\cO_n$ and $\Gamma_k$ in the fluctuation fields are readily computed using the xPert extension \citep{Brizuela:2008ra} of xAct. For completeness, the relevant expressions are listed in Table \ref{Tab.2}. The $d$-dependent coefficients $C_i$ multiplying the curvature terms in $\Gamma_k^{(2)}$ are
\be\label{Cdefs}
C_T = \frac{d^2-3d+4}{d(d-1)} \, , \quad 
C_V = \frac{d-2}{d} \, , \quad
C_S = \frac{d-4}{2(d-1)} \, , \quad
C_\sigma = - (d-2) \, . 
\ee
\begin{table}[h!]
	\begin{center}
		\begin{tabular}{ll}
		operator & value of the matrix element	\\ \hline \hline
		$\bigg.  \cO_n^{(2)}\big|_{h^{\rm T}h^{\rm T}}$ \qquad & $- \frac{1}{2} \, \Rb^{n-1} \,  \left[ n  \Delta - \left( \frac{2n(d-2)}{d(d-1)} - 1 \right) \Rb \right]$ \\[1.5ex]
			$\bigg.  \cO_n^{(2)}\big|_{\xi\xi}$ & $\frac{2n-d}{2d} \, \Rb^n$ \\[1.5ex]
		$\Big.  \cO_n^{(2)}\big|_{hh}$ & 
		$ \frac{n(n-1)(d-1)^2}{d^2} \, \Rb^{n-2} \Delta^2
		+ \frac{n(d^2 - (4n-1)(d-1)-1 )}{2d^2} \, \Rb^{n-1} \Delta + 
		\left( \frac{d-2}{4d} - \frac{n(d-n-1)}{d^2} \right) \Rb^n $
		\\[1.5ex]
				$\Big.  \cO_n^{(2)}\big|_{\sigma\sigma}$ & 
				$\frac{n(n-1)(d-1)}{d} \Rb^{n-2} \, \Delta^2 + \frac{n (d - 2 n)}{2d}  \Rb^{n-1} \, \Delta - \frac{d-2n}{2d} \, \Rb^n $ \\[1.5ex]
						$\bigg.  \cO_n^{(2)}\big|_{\sigma h}$ & 
						$\Big[ \frac{ n (n-1)(d-1)}{d} \Rb^{n-2} \Delta + \frac{n(d-2n)}{2d} \Rb^{n-1}  \Big] 
						\Big[ \frac{d-1}{d} \Delta^2 - \frac{1}{d} \Rb \Delta \Big]^{1/2}$ \\ \hline
$\bigg. \Gamma^{(2)}_k\big|_{h^{\rm T}h^{\rm T}}$  & $\frac{1}{32 \pi G_k} \left[\Delta - 2 \Lambda_k + C_T \Rb \right]$ \\[1.5ex]
$\Big. \Gamma^{(2)}_k\big|_{\xi\xi}$ & $\frac{1}{32 \pi G_k \, \alpha} \left[ \Delta - \frac{1}{d} \Rb + \alpha \, (C_V \Rb - 2 \Lambda_k) \right]$  \\[1.5ex]
$\Big. \Gamma^{(2)}_k\big|_{hh}$ & $\frac{1}{32 \pi G_k \, \alpha \, d^2} \left[ 2 \rho^2 \Delta - \alpha (d-1)(d-2)\, (\Delta - \frac{d}{d-1} \Lambda_k + C_S \Rb) \right]$  \\[1.5ex]
$\Big. \Gamma^{(2)}_k\big|_{\sigma\sigma}$ & $\frac{1}{32 \pi G_k \, \alpha \, d}
\left[ 2(d-1) \Delta - 2 \Rb - \alpha \left((d-2) \Delta + 2 d \Lambda_k + C_\sigma \Rb \right) \right]$  \\[1.5ex]
$\Big. \Gamma^{(2)}_k\big|_{\sigma h}$  & $\frac{2\rho + (d-2) \alpha}{32 \pi G_k \, \alpha \, d}
\left[ \Delta \left( \frac{d-1}{d} \Delta - \frac{1}{d} \Rb \right) \right]^{1/2}$ \\[1.5ex]
$\Big.\Gamma^{(2)}_k\big|_{\Cb^{\rm T}C^{\rm T}}$  &  $\left[ \Delta - \frac{1}{d} \Rb \right]$ \\[1.5ex]
$\bigg. \Gamma^{(2)}_k\big|_{\bar{\eta}\eta}$ &  $\frac{2}{d} \left[ (d-1-\rho) \Delta - \Rb \right]$ 
\\ \hline						
	$\bigg. \cR_k \big|_{h^{\rm T}h^{\rm T}}$   & $\frac{1}{32 \pi G_k} \, R_k$ \\[1.5ex]
	$\Big. \cR_k \big|_{\xi\xi}$ & $\frac{1}{32 \pi G_k \, \alpha} R_k$  \\[1.5ex]
	$\Big. \cR_k \big|_{hh}$ & $\frac{1}{32 \pi G_k \, \alpha \, d^2} \, \left[2 \rho^2 - \alpha (d-1)(d-2) \right] \, R_k$ \\[1.5ex]
	$\Big. \cR_k \big|_{\sigma\sigma}$ & $\frac{1}{32 \pi G_k \, \alpha \, d} \left[2(d-1) - \alpha \, (d-2) \right] \, R_k$  \\[1.5ex]
	$\Big. \cR_k \big|_{\sigma h}$ & $\frac{2\rho + (d-2) \alpha}{32 \pi G_k \, \alpha \, d} \left\{ 
							\left[ P_k \left( \tfrac{d-1}{d} P_k - \tfrac{1}{d} \Rb \right) \right]^{1/2}  -
							\left[ \Delta \left( \tfrac{d-1}{d} \Delta - \tfrac{1}{d} \Rb \right) \right]^{1/2}  \right\}$ \\[1.5ex]
	$\Big. \cR_k \big|_{\bar{C}^{\rm T} C^{\rm T}}$ & $R_k$ \\[1.5ex]
	$\bigg. \cR_k \big|_{\bar{\eta}\eta}$ & $ \frac{2}{d} \, \left[d-1-\rho \right] \, R_k$ 		
						\\[1.2ex] 		   \hline \hline 
		\end{tabular}
		\caption{\label{Tab.2} Components of the Hessians entering the right-hand side of the composite operator equation \eqref{CompositeMaster2} and the Wetterich equation evaluated for the Einstein-Hilbert truncation. The fluctuations are expressed by the component fields \eqref{TTdec} and \eqref{Tdec} followed by the field redefinitions \eqref{fieldredef} and \eqref{fredghost}. The matrix elements are labeled by the fluctuation fields, i.e., $\cO_n^{(2)}|_{h^{\rm T}h^{\rm T}}$ results from taking two functional derivatives of $\cO_n$ with respect to $h^{\rm T}$. The off-diagonal terms are symmetric, e.g., $\cO_n^{(2)}|_{h \sigma} = \cO_n^{(2)}|_{\sigma h}$.}
	\end{center}
\end{table}
\newpage


\section*{Acknowledgments}
F.S.\ thanks T.\ Budd, L.\ Lionni, C.\ Pagani, and M.\ Reuter for inspiring discussions. Furthermore, we are grateful to W.\ Houthoff for participating in the earlier parts of the program. The work of F.S.\ is supported by the Netherlands Organisation  for  Scientific  Research  (NWO)  within  the  Foundation  for  Fundamental Research on Matter (FOM) grant 13VP12. A.K.\ is supported by the Foundation for Theoretical Physics Development Basis and by the RFBR grant No. 20-02-00297.

\bibliographystyle{frontiersinHLTH&FPHY} 





\end{document}